\documentclass[prd,twocolumn,superscriptaddress,preprintnumbers,amsmath,amssymb,final]{revtex4-1}
\usepackage{bm}
\usepackage[colorlinks=true,linkcolor=blue,citecolor=blue]{hyperref}
\usepackage{amsthm}
\usepackage{amsfonts}
\usepackage{enumerate}
\usepackage{latexsym}
\usepackage{ifpdf}
\usepackage{graphicx}
\usepackage{makeidx}
\expandafter\ifx\csname package@font\endcsname\relax\else
 \expandafter\expandafter
 \expandafter\usepackage
 \expandafter\expandafter
 \expandafter{\csname package@font\endcsname}
 \fi
\usepackage{color}
\usepackage{times}

\begin{document}

\title{Unconventional crystal field splitting in non-centrosymmetric BaTiO$_3$ thin films}

\author{Yang~Song}
\affiliation{Ningbo Institute of Materials Technology and Engineering, Chinese Academy of Sciences, Ningbo, Zhejiang 315201, China}
\affiliation{Center of Materials Science and Optoelectronics Engineering, University of Chinese Academy of Sciences, Beijing 100049, China}
\author{Xiaoran~Liu}
\affiliation{Department of Physics and Astronomy, Rutgers University, Piscataway, New Jersey 08854, USA}
\author{Fangdi~Wen}
\affiliation{Department of Physics and Astronomy, Rutgers University, Piscataway, New Jersey 08854, USA}
\author{M.~Kareev}
\affiliation{Department of Physics and Astronomy, Rutgers University, Piscataway, New Jersey 08854, USA}
\author{Ruyi~Zhang}
\affiliation{Ningbo Institute of Materials Technology and Engineering, Chinese Academy of Sciences, Ningbo, Zhejiang 315201, China}
\affiliation{Center of Materials Science and Optoelectronics Engineering, University of Chinese Academy of Sciences, Beijing 100049, China}
\author{Yujuan~Pei}
\affiliation{Ningbo Institute of Materials Technology and Engineering, Chinese Academy of Sciences, Ningbo, Zhejiang 315201, China}
\affiliation{Center of Materials Science and Optoelectronics Engineering, University of Chinese Academy of Sciences, Beijing 100049, China}
\author{Jiachang~Bi}
\affiliation{Ningbo Institute of Materials Technology and Engineering, Chinese Academy of Sciences, Ningbo, Zhejiang 315201, China}
\affiliation{Center of Materials Science and Optoelectronics Engineering, University of Chinese Academy of Sciences, Beijing 100049, China}
\author{Padraic Shafer}
\affiliation{Advanced Light Source, Lawrence Berkeley National Laboratory, Berkeley, CA 94720, USA}
\author{Alpha~T.~N'Diaye}
\affiliation{Advanced Light Source, Lawrence Berkeley National Laboratory, Berkeley, CA 94720, USA}
\author{Elke~Arenholz}
\affiliation{Advanced Light Source, Lawrence Berkeley National Laboratory, Berkeley, CA 94720, USA}
\author{Se Young Park}
\email{sp2829@snu.ac.kr}
\affiliation{Center for Correlated Electron Systems, Institute for Basic Science, Seoul 08826, Korea}
\affiliation{Department of Physics and Astronomy, Seoul National University, Seoul 08826, Korea}
\author{Yanwei~Cao}
\email{ywcao@nimte.ac.cn}
\affiliation{Ningbo Institute of Materials Technology and Engineering, Chinese Academy of Sciences, Ningbo, Zhejiang 315201, China}
\affiliation{Center of Materials Science and Optoelectronics Engineering, University of Chinese Academy of Sciences, Beijing 100049, China}
\author{J.~Chakhalian}
\affiliation{Department of Physics and Astronomy, Rutgers University, Piscataway, New Jersey 08854, USA}

\date{\today}

\begin{abstract} 

Understanding the crystal field splitting and orbital polarization in non-centrosymmetric systems such as ferroelectric materials is fundamentally important. In this study, taking BaTiO$_3$ as a representative material we investigate titanium crystal field splitting and orbital polarization in non-centrosymmetric TiO$_6$ octahedra with resonant X-ray linear dichroism at Ti $L_{2,3}$-edge. The high-quality BaTiO$_3$ thin films were deposited on DyScO$_3$ (110) single crystal substrates in a layer-by-layer way by pulsed laser deposition. The reflection high-energy electron diffraction and element specific X-ray absorption spectroscopy were performed to characterize the structural and electronic properties of the films. In sharp contrast to conventional crystal field splitting and orbital configuration ($d_{xz}$/$d_{yz}$ $<$ $d_{xy}$ $<$ $d_{3z^2-r^2}$ $<$ $d_{x^2-y^2}$ ~ or ~ $d_{xy}$ $<$ $d_{xz}$/$d_{yz}$ $<$ $d_{x^2-y^2}$ $<$ $d_{3z^2-r^2}$) according to Jahn-Teller effect, it is revealed that $d_{xz}$, $d_{yz}$, and $d_{xy}$ orbitals are nearly degenerate, whereas $d_{3z^2-r^2}$ and $d_{x^2-y^2}$ orbitals are split with an energy gap $\sim$ 100 meV in the epitaxial BaTiO$_3$ films. The unexpected degenerate states $d_{xz}$/$d_{yz}$/$d_{xy}$ are coupled to Ti-O displacements resulting from competition between polar and Jahn-Teller distortions in non-centrosymmetric TiO$_6$ octhedra of BTO films. Our results provide a route to manipulate orbital degree of freedom by switching electric polarization in ferroelectric materials.

\end{abstract}
 
%\pacs{}
%\keywords{}
\maketitle
\newpage

\section{Introduction}
Ferroelectric materials (oxide films in particular) which exhibit robust spontaneous electric polarization that can be reoriented with an external electric field, have attracted attention because of their extensive applications such as transistors, memories, and high frequency devices realized in various systems \cite{NRM-2016-Martin,RMP-2005-Dawber,Nature-1992-Cohen,RMP-1994-Resta,ARMR-2007-Schlom,JAP-2006-Setter,Science-2005-NS,NL-2002-WY,NM-2007-Ramesh,Nature-2006-Eer}. Generally, this spontaneous polarization is coupled to lattice structures such as Ti-O displacements in BaTiO$_3$ (BTO), a typical ferroelectric material with non-centrosymmetric TiO$_6$ octahedra \cite{Nature-2005-HN,PNAS-2017-HG,NC-2018-YC,PRB-2019-Cordero}. Due to crystal field splitting being sensitive to lattice distortions, the ion displacements in ferroelectric materials can modify the crystal field splitting, the study of which is fundamentally important for understanding the properties of ferroelectric materials and non-centrosymmetric superconductors, \textit{e.g.}, the coexistence of superconducting and ferroelectric states in doped perovskite oxides \cite{NatPhys-2017-CR, Natphys-2017-MG}, however, crystal field splittings in these system have not been extensively studied yet.  \cite{NC-2013-CS,PRB-2010-EA}.

\begin{figure*}[]
	\includegraphics[width=0.8\textwidth]{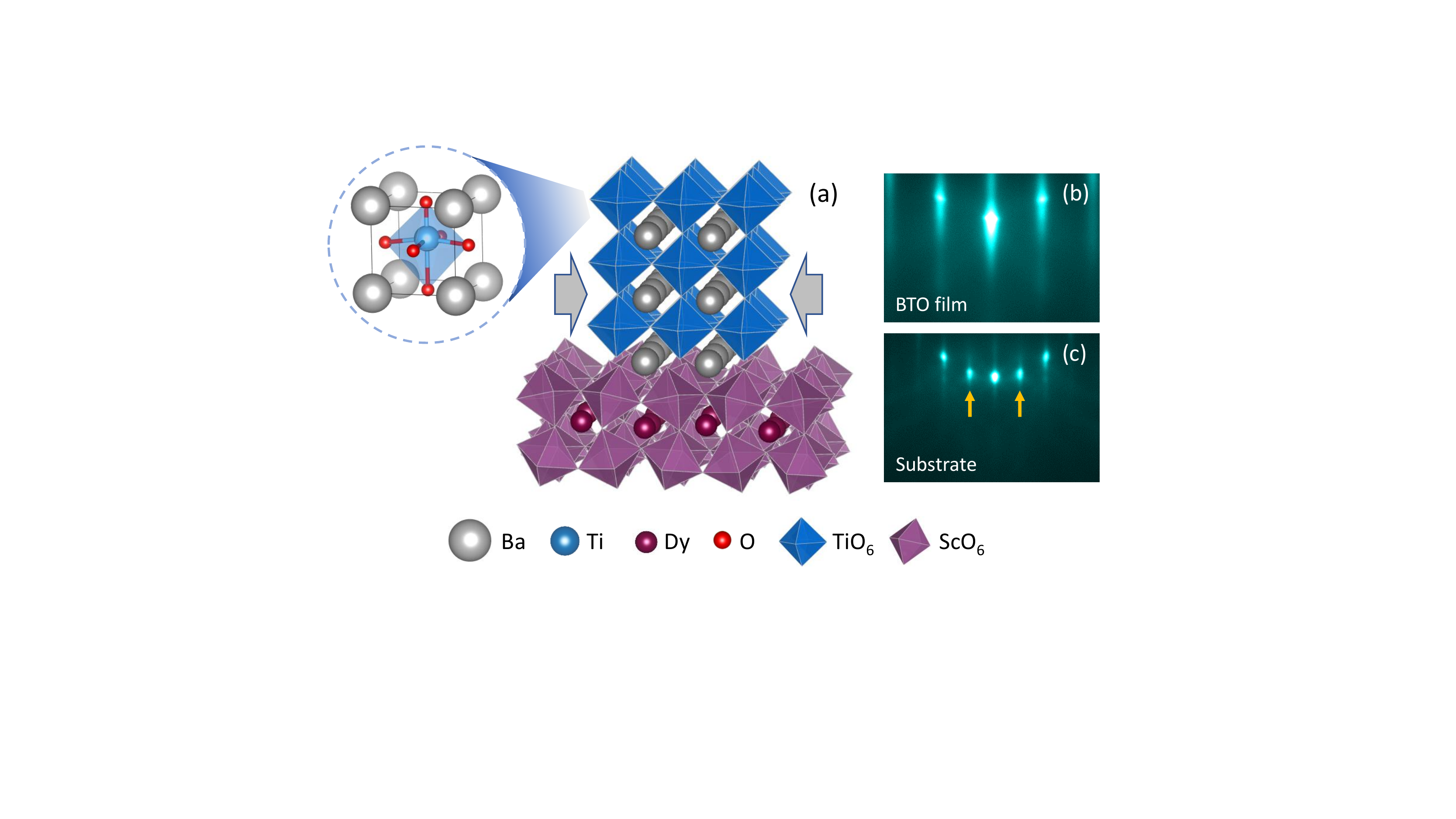}
	\caption{\label{fig1} (a) Schematic of epitaxial BTO film on a DSO substrate. The two arrows indicate biaxial compressive strain on BTO film. (b-c) RHEED patterns of (b) BTO film (after growth at room temperature) and (c) DSO (110) substrate (before growth). The yellow arrows indicate the half-order Bragg peaks.}
\end{figure*}

\begin{figure*}[]
	\includegraphics[width=0.8\textwidth]{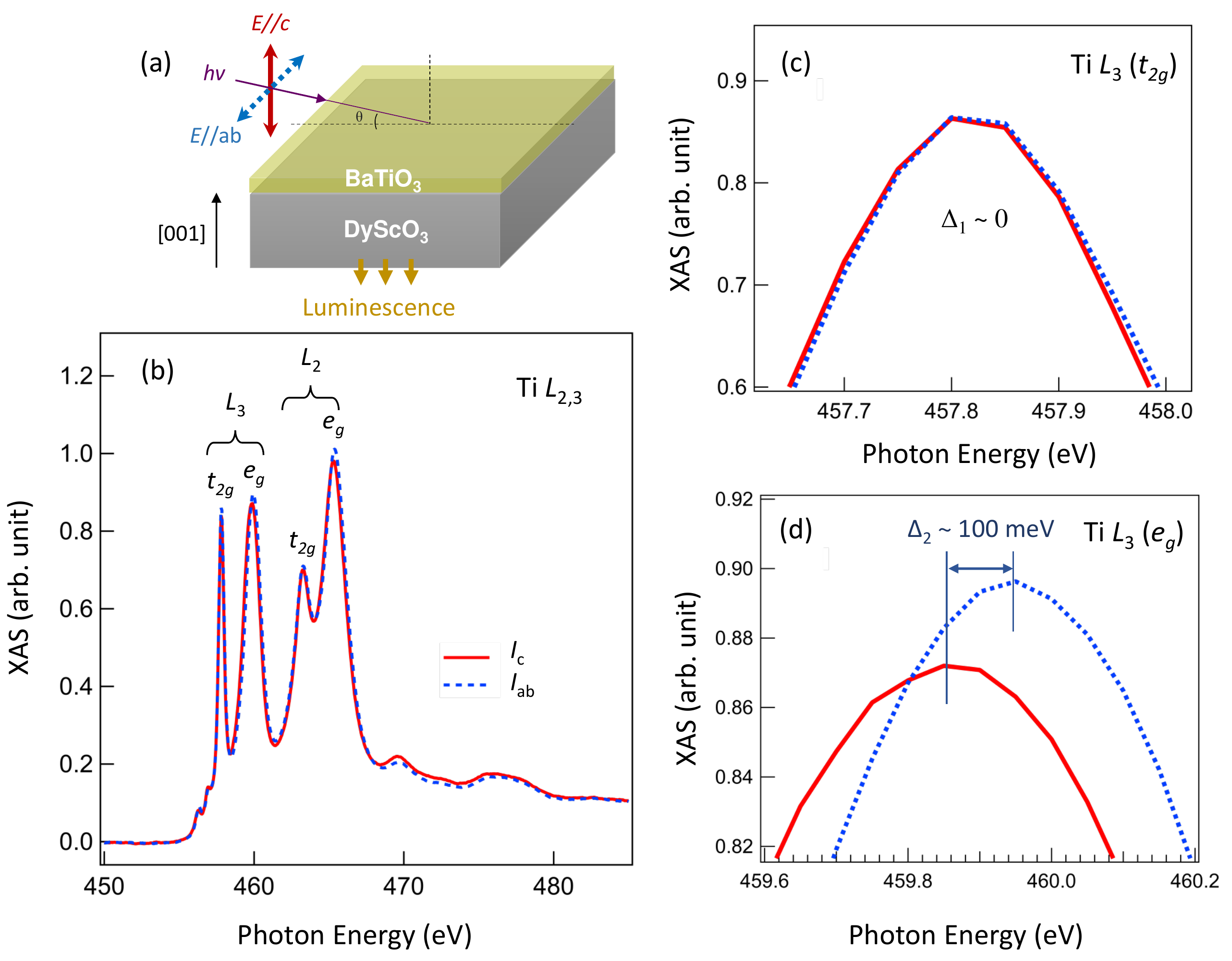}
	\caption{\label{fig2} (a) Schematic of experimental setup. \textit{E} $\parallel$ $c$ and  \textit{E} $\parallel$ $ab$ ( \textit{E} is the polarization vector of the photon) indicate out-of-plane (red solid line, $I_\textrm{c}$) and in-plane (blue dash line, $I_\textrm{ab}$) linearly polarized incident X-ray, respectively. The grazing angle $\theta$ = $20^{\circ}$. The yellow arrows at the bottom of DSO substrates indicates the bulk-sensitive luminescence yield detection mode. The arrow along [001] direction (pseudocubic notation) indicates the growth direction of BTO films. (b) XAS of BTO films at Ti $L_{2,3}$-edge at room temperature. All collected spectra are repetitively measured more than six times. Enlarged XAS spectra at (c) Ti $L_3$ ($t_{2g}$) and (d) Ti $L_3$ ($e_g$) absorption peaks. $\Delta_{1}$ and $\Delta_{2}$ are defined as the splitting between $d_{xy}$ and $d_{xz}$/$d_{yz}$ orbitals, and between $d_{x^2-y^2}$ and $d_{3z^2-r^2}$ orbitals, respectively}
\end{figure*}

To address the above concern, we take BTO thin films as a representative material to investigate crystal field splitting and orbital polarization in the presence of polar distortions induced by ordering of ferroelectric dipoles. Bulk BaTi$^{4+}$O$_3$ with 3\textit{d}$^{0}$ electron configuration undergoes complex structural and ferroelectric phase transitions upon cooling, \textit{e.g.}, from high temperature cubic to tetragonal (393~K), tetragonal to orthorhombic (278~K), and orthorhombic to rhombohedral (183~K) , where ferroelectric properties are present below 393~K \cite{JPC-1993-Kw,PRL-1997-Ish,JPCM-2002-Hay}. Moreover, under biaxial compressive strain (such as epitaxial tetragonal BTO films on DyScO$_3$ substrates) the transition temperature can be enhanced to nearly 500~K and the remnant polarization is at least 2.5 times higher than bulk BTO single crystals \cite{Science-2004-Choi}. Due to strong $B$-site ferroelectricity at room temperature \cite{JACS-2008-MS,PRB-2018-CM,PRL-1994-PB,APL-1996-AK,PRB-2017-Jin}, the BTO single crystals are widely used as ferroelectric substrates for epitaxial thin film synthesis \cite{APL-2000-Emo,APL-2011-Brivio,PRB-2012-Alberca}, where its epitaxial films can be employed to manipulate interfacial electric field for controlling order parameters and achieving novel functionalities \cite{NM-2018-LW,AFM-2016-BC,Sci-rep-2017-Takahashi}.

In this paper, high-quality tetragonal BTO thin films were grown on DyScO$_3$ (DSO) (110) single crystal substrates in a layer-by-layer way by pulsed laser deposition (PLD). The structural and electronic properties were characterized by reflection high-energy electron diffraction (RHEED) and element-specific X-ray absorption spectroscopy (XAS). In sharp contrast to conventional crystal field splitting and orbital polarization from the Jahn-Teller effect, an anomalous orbital structure, nearly degenerate $t_{2g}$ ($d_{xz}$/$d_{yz}$/$d_{xy}$) and split $e_{g}$ ($d_{3z^2-r^2}$ $<$ $d_{x^2-y^2}$ about 100 meV) orbitals, was revealed in epitaxial BTO films, resulting from the competition between polar and Jahn-Teller distortions in non-centrosymmetric TiO$_6$ octhedra of polar BTO films.

\section{Experiments and first-principles calculations}
As shown in Fig.~1, the BTO films (20 unit cells, $\sim$ 8.2 nm) had been grown along [110] (orthorhombic notation, corresponding to [001] orientation in a pseudocubic notation) DSO substrates (5 $\times$ 5 $\times$ 0.5 mm$^3$) by PLD, using a KrF excimer laser operating at \textit{$\lambda$}~=~248~nm and 2 Hz pulse rate with 2~J/cm$^2$ fluence. During the growth, the oxygen pressure was kept at $\sim$ 10$^{-6}$ Torr, the temperature of the substrates was $\sim$~$850\,^{\circ}{\rm C}$ (from reader of infrared pyrometers). At room temperature, the bulk lattice parameters are \textit{a} = 3.99~\AA~ and \textit{c} = 4.04~\AA~for tetragonal BTO,  and \textit{a} = 3.95~\AA ~for DSO (pseudo cubic). Due to the lattice mismatch, the epitaxial BTO films on DSO are under biaxial compressive strain, indicated by a pair of arrows in Fig.~1(a). In order to monitor the growth of the BTO thin film, an ${in}$-${situ}$ RHEED was performed during the deposition. The sharp RHEED patterns, Fig.~1(b) and (c), suggest a high-quality two-dimensional growth of the BTO films. In order to investigate the electronic structure of BTO films, linearly polarized XAS in luminescence yield detection mode (bulk-sensitive, see Fig. 2(a)) was performed at room temperature at beamline 4.0.2 of the Advanced Light Source (ALS, Lawrence Berkeley National Laboratory) \cite{APL_2015_Cao}, and the preliminary data were collected at beamline 6.3.1. We have carried out first-principles density-functional theory (DFT) calculations within the local density approximation (LDA) \cite{PRL-1980-Cep,PRB-1981-Per}. The calculations are performed using the Vienna \textit{ab-initio} simulation package (VASP) \cite{PRB-1996-Kre,PRB-1999-Kre}. The projector augmented wave (PAW) \cite{PRB-1994-Blo} is used with an energy cut-off of 600 eV. The Brillouin zone is sampled with a $8\times 8\times 8$ $k$-point grid for 5-atom unit cell of BTO. Convergence is reached if the consecutive energy difference is less than 10$^{-6}$ eV for electronic iterations and 10$^{-5}$ eV for ionic relaxations. The polarization is calculated using the Berry-phase method \cite{PRB-1993-Kin} as implemented in VASP. The tight-binding parameters for Ti $d$-orbitals are calculated by Wannier90 package \cite{CPC-2014-Mos}. 

\section{Results and discussions}
Figure~2(b) shows the XAS spectra of BTO films at Ti $L_{2,3}$-edge. There are four well split characteristic peaks, arising from the excitations from Ti $2p$ to Ti 3$d$ states (the electronic configuration changes from Ti $2p^63d^0$ to $2p^53d^1$). Generally, in octahedral symmetry, the transition-metal $d$ bands split into $t_{2g}$ ($d_{xy}$, $d_{xz}$, $d_{yz}$) and $e_g$ ($d_{3z^2-r^2}$, $d_{x^2-y^2}$) \cite{PRL-2009-Sal,AM-2013-Sal,PRL-2013-Sal,PRL-2014-Pes,Nmat-2013-Lee,PRL-2016-Cao,npj-2016-Cao}. The degenerate $t_{2g}$ and $e_{g}$ bands further split when the octahedron experiences uniaxial elongation or compression \cite{PRB-1997-DL}, along the c-axis where the Jahn-Teller distortion leads to orbital structure $d_{xz}$/$d_{yz}$ $<$ $d_{xy}$ $<$ $d_{3z^2-r^2}$ $<$ $d_{x^2-y^2}$ ~ or ~ $d_{xy}$ $<$ $d_{xz}$/$d_{yz}$ $<$ $d_{x^2-y^2}$ $<$ $d_{3z^2-r^2}$, respectively. Using that intensity of linearly polarized XAS carries the information of crystal field splitting and orbital polarization \cite{PRL-2009-Sal,AM-2013-Sal,PRL-2013-Sal,PRL-2014-Pes,Nmat-2013-Lee,PRL-2016-Cao,npj-2016-Cao},  Fig. 2(c) and (d) show the differences between out-of-plane ($I_\textrm{c}$) and in-plane ($I_\textrm{ab}$) polarized XAS. As seen in the figure, the $t_{2g}$ orbitals are nearly degenerate ($\Delta_{1}$ $\sim$ 0, see Fig.~2(c)) , whereas the splitting $\Delta_{2}$ between $d_{x^2-y^2}$ and $d_{3z^2-r^2}$ orbitals in $e_{g}$ state is about 100 meV, shown in Fig.~2(d). The same orbital spitting is observed at Ti $L_{2}$-edge as well. This kind of orbital structure in BTO films ( $d_{xz}$/$d_{yz}$/$d_{xy}$ $<$ $d_{3z^2-r^2}$ $<$ $d_{x^2-y^2}$) is unexpected and in sharp contrast to conventional crystal field splitting ( $d_{xz}$/$d_{yz}$ $<$ $d_{xy}$ $<$ $d_{3z^2-r^2}$ $<$ $d_{x^2-y^2}$) in elongated TiO$_6$ octahedra along the $c$ axis \cite{APL-2016-YC}. 

To further understand this anomalous orbital structure in BTO films, we obtain the spectra of X-ray linear dichroism (XLD), which is defined as XLD = $I_\textrm{c}$ - $I_\textrm{ab}$ in this work. Figure 3 shows the XLD spectra of the biaxial compressive BTO and CaTiO$_3$ (CTO) films at Ti $L_{2,3}$-edge. We note that among $A$TiO$_3$ titanates ($A$ = Ba, Sr, Ca) only \textit{without} Ti-O displacements from the center of TiO$_6$ octhedra are taken as a reference material to compare with ferroelectric BTO films \textit{with} Ti-O displacements. As highlighted in the orange area in Fig.~3, the XLD of CTO indicates crystal field splitting ($d_{xz}$/$d_{yz}$ $<$ $d_{xy}$) within $t_{2g}$ states, whereas $t_{2g}$ orbitals are nearly degenerate ($d_{xz}$/$d_{yz}$/$d_{xy}$) in BTO films. 

\begin{figure}
	\includegraphics[width=0.45\textwidth]{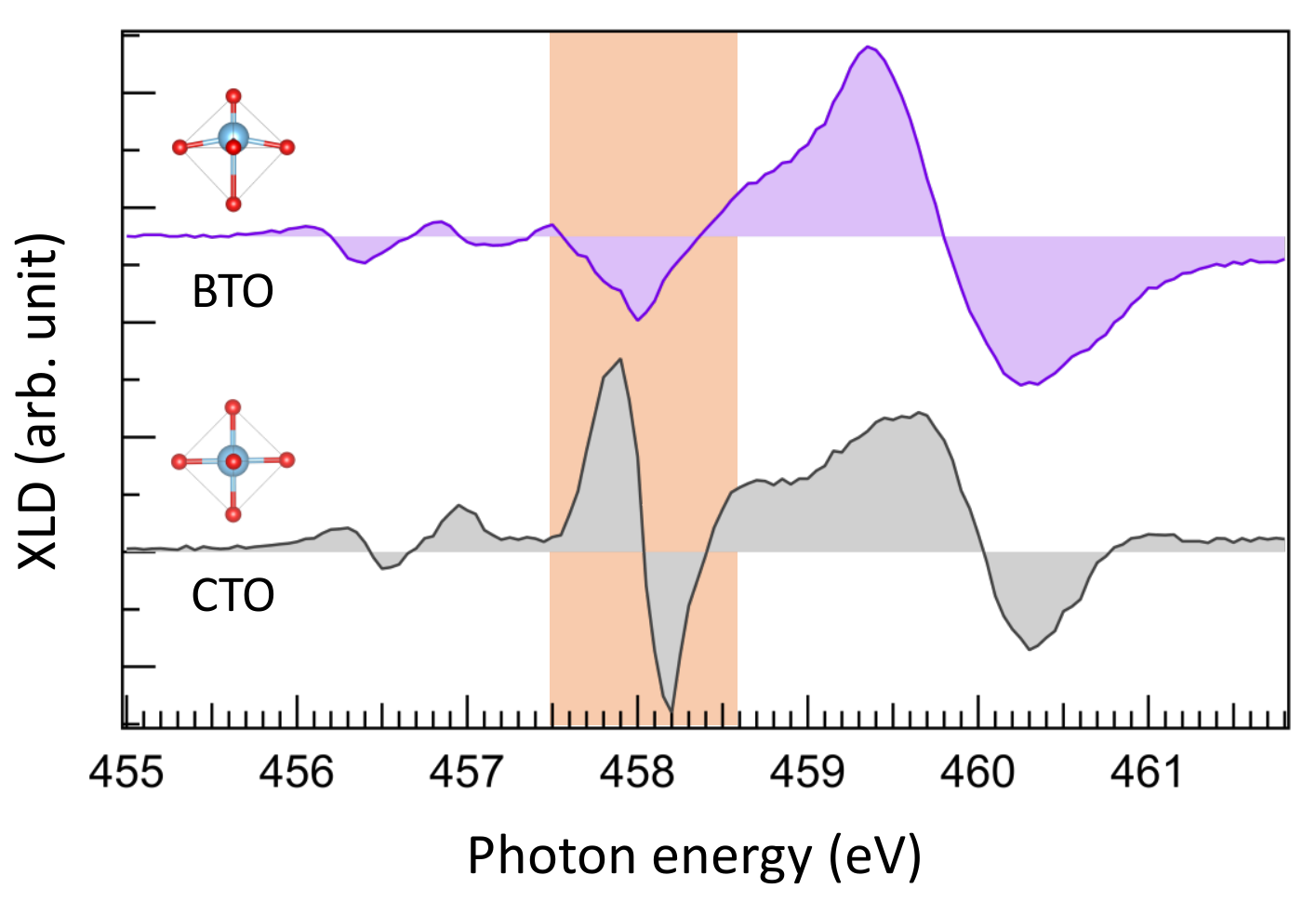}
	\caption{\label{fig3} XLD of BTO (in purple, with Ti-O displacements) and CTO (in gray, without Ti-O displacements) \cite{APL-2016-YC} thin films at room temperature. The rectangular orange shadow highlights the contribution of ferroelectric polarization on crystal field splitting and orbital polarization.}
\end{figure}

The discrepancy between the degenerate $t_{2g}$ orbitals and the expected crystal field splitting from the compressive strain strongly suggest an additional mechanism behind unusual orbital splitting. Since the bulk BTO is ferroelectric at room temperature \cite{JPC-1993-Kw}, it is natural to include the polar distortion in BTO and investigate its contribution to the orbital splitting. Using the first-principles DFT method, we first calculate the effect of the epitaxial strain on the Jahn-Teller distortion (or change in the $c/a$ ratio) without considering the polar distortion. With 1.25\% compressive strain based on the lattice mismatch between BTO and DSO, the $c/a$ ratio obtained by strained-bulk calculation increases to 1.02. Next, we allow a polarization along the $c$-axis, experimentally observed under the compressive strain \cite{PRB-2004-Di}. The relaxed atomic structure with 1.25\% compressive epitaxial strain lowers the total energy by 19 meV per formula unit from a polar distortion, mainly contributed by Ti displacement from the center of the octahedron with Ti-O-Ti angle of 172$^{\circ}$. The calculated polarization is 33 $\mu$C/cm$^{2}$ along the c axis with further increase in the $c/a$ ratio to 1.04. 

\begin{table*}
\caption{\label{tab1} On-site energies and major hopping parameters of Ti $d$-orbital for BTO (in eV) obtained from Wannier functions of Ti $d$-band. For hopping parameter $t$, $||$ and $\perp$ represent the in-plane and out-of-plane hopping, respectively.}
\begin{ruledtabular}
\begin{tabular}{ccc|cccccc}
& \multicolumn{2}{c|}{On-site energy difference} &  \multicolumn{5}{c}{Intra orbital hopping}\\
  & $E_{xz/yz}-E_{xy}$ & $E_{3z^{2}-r^{2}}-E_{x^2-y^2}$ & $t^{||}_{xy}$ & $t^{||}_{xz/yz}$ &  $t^{\perp}_{xz/yz}$ & $t^{||}_{x^2-y^2}$ & $t^{\perp}_{3z^{2}-r^{2}}$  \\ \hline
Cubic & 0  & 0 & -0.31 & -0.31 & -0.31 & -0.50 & -0.66 \\
Tetragonal ($P=0$)&  -0.13 & -0.45 & -0.30   & -0.33 & -0.25& -0.50& -0.64 \\
Cubic ($P \neq 0$) &0.15 & 0.13  & -0.30  & -0.26 & -0.28 & -0.48 & -0.66 \\
\end{tabular}
\end{ruledtabular}
\end{table*}

\begin{figure}[]
	\includegraphics[width=0.45\textwidth]{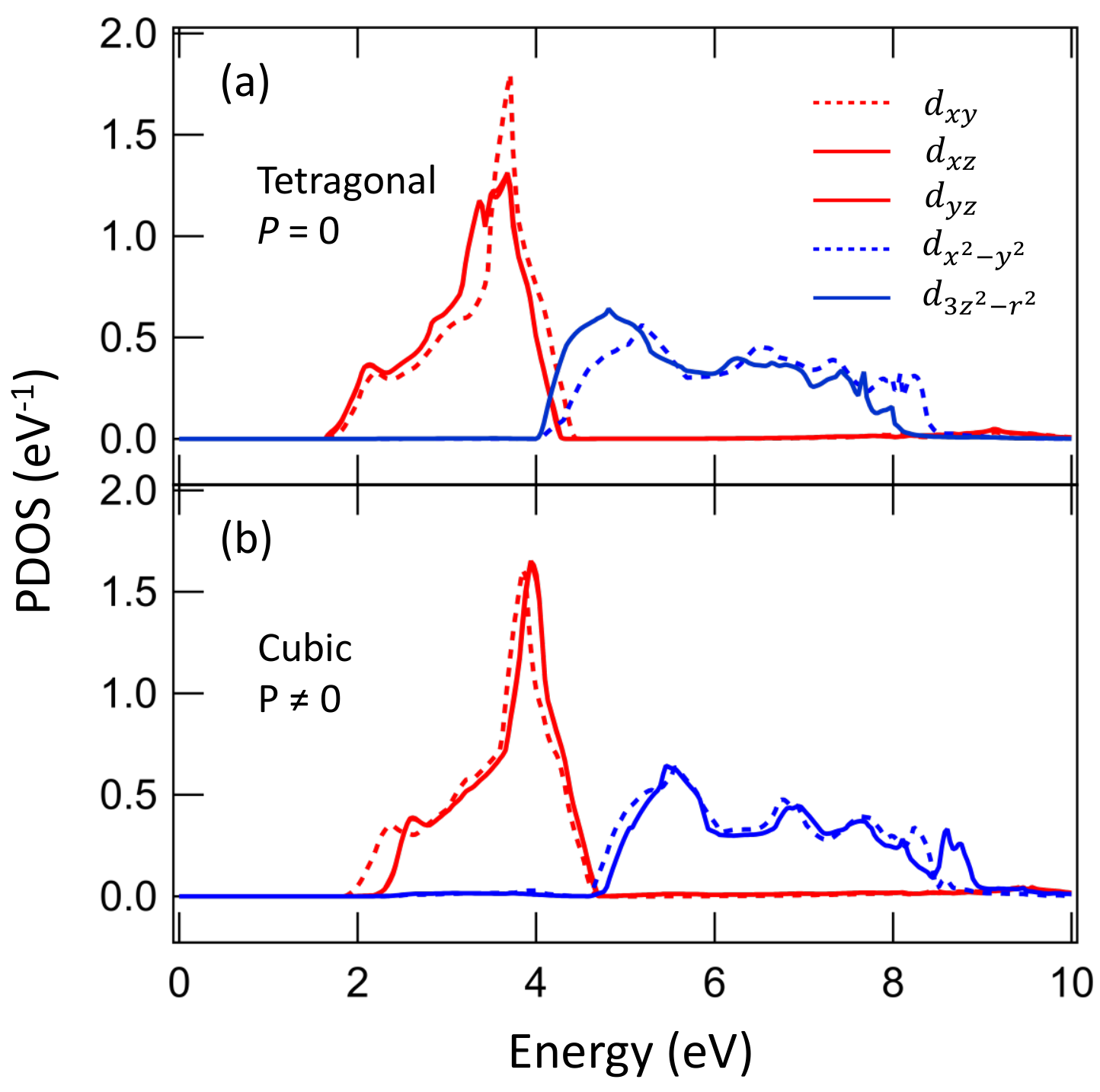}
	\caption{\label{fig4} PDOS of the strained bulk calculations with 1.25\% biaxial compressive strain. PDOS of (a) tetragonal BTO with zero polarizations ($c/a=1.04$) and (b) cubic BTO with polar distortion.}
\end{figure}

In order to investigate the effect of the Jahn-Teller and polar distortion to orbital polarization, we consider two atomic configurations: one with Jahn-Teller distortion with $c/a = 1.04$ with zero polar distortion and the other with cubic lattice constant of 1.25\% compressive strain for all the lattice constant including polar distortion of the strained bulk calculation. The projected density of states (PDOS) and Wannier tight-binding parameters are presented in Fig.~\ref{fig4} and table~\ref{tab1}, respectively. In Fig. 4(a), we find that the effect of the Jahn-Teller distortion from the compressive strain is straightforwardly shown in the PDOS with $d_{xz}$/$d_{yz}$ $<$ $d_{xy}$ and  $d_{3z^2-r^2}$ $<$ $d_{x^2-y^2}$ orbital splitting, consistent with on-site energy difference $E_{xz/yz} - E_{xy}$ = $-0.13$ eV for $t_{2g}$ orbitals and  $E_{3z^2-r^2} - E_{x^2-y^2}$ = $-0.45$ eV for $e_{g}$ orbitals (Table~\ref{tab1}).  We note that the splitting between $t_{2g}$ orbitals is smaller than that of $e_{g}$ orbitals due to the relatively weaker $\pi$ bonding of $t_{2g}$ orbitals compared to the $\sigma$ bonding of $e_{g}$. In contrast, the inclusion of the polar distortion without considering the Jahn-Teller distortion results in the opposite trend in the PDOS as shown in Fig.~\ref{fig4}(b):  $d_{xz}$/$d_{yz}$ $>$ $d_{xy}$ and  $d_{3z^2-r^2}$ $>$ $d_{x^2-y^2}$ orbital splitting, consistent with the on-site energy difference of $E_{xz/yz} - E_{xy}$ = 0.15 eV for $t_{2g}$ orbitals and of $E_{3z^2-r^2} - E_{x^2-y^2}$ = 0.13 eV for $e_{g}$ orbitals (Table~\ref{tab1}). We note that the splitting between the $t_{2g}$ orbitals from the polar distortion is larger than that between $e_{g}$ orbitals, consistent with relatively larger reduction in the in-plane hopping $t^{||}_{xz/yz}$ about 50 meV than that of $t^{||}_{x^{2}-y^{2}}$ about 20 meV compared with the cubic hopping parameters. This  suggests that the bonding between the $t_{2g}$ orbitals is more sensitive to the Ti-O-Ti angle than that between the $e_{g}$ orbitals. 

%We find that the in-plane hopping between $d_{xy}$ orbitals ($t^{||}_{xy}$) and out-of-plane hopping between $d_{3z^{2}-r^{2}}$ orbitals ($t^{\perp}_{3z^{2}-r^{2}}$) are insensitive to the polar distortion.

\begin{figure}[]
	\includegraphics[width=0.45\textwidth]{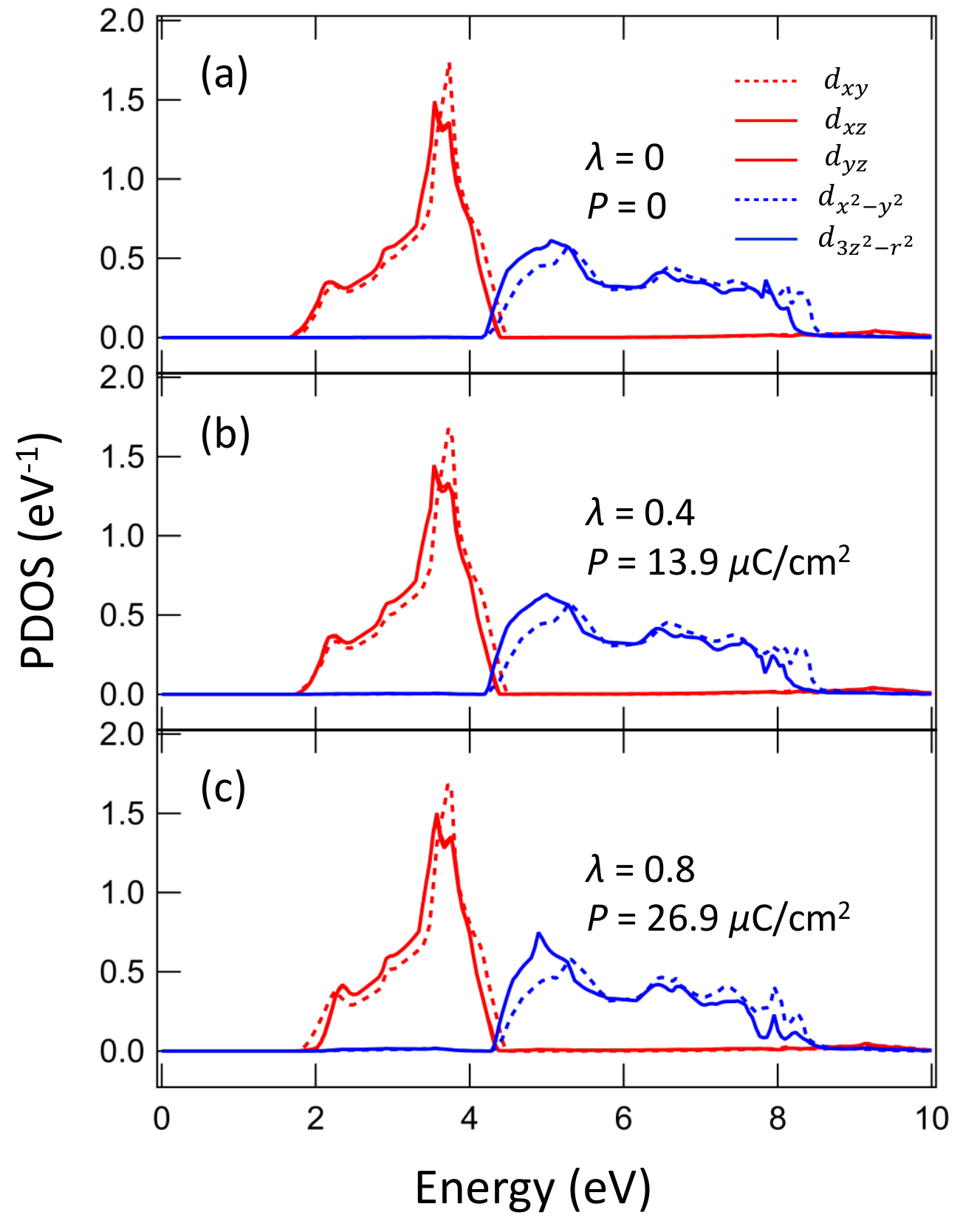}
	\caption{\label{fig5} PDOS of the atomic structures linear interpolated between non-polar strained bulk BTO structure with $c/a=1.02$ and polar strained bulk BTO  structure with $P=33\; \mu C$/cm$^{2}$. The symbol $\lambda$ is defined as the parameter of the linear interpolation ($\lambda=0$ and 1 for non-polar and polar structures. PDOS for (a) $\lambda$ = 0, (b) $\lambda$ = 0.4 ($P$ = 14 $\mu$C/cm$^{2}$) and (c) $\lambda$ = 0.8 ($P$ = 27 $\mu$C/cm$^{2}$).
	}
\end{figure}

From the opposite tendencies of the orbital splitting between the Jahn-Teller and polar distortions, we can understand the orbital polarization observed in the XLD measurements; the degeneracy of the $t_{2g}$ orbitals is maintained by the cancellation of orbital splitting from Jahn-Teller and polar distortions which are similar in amount ($-0.13$ eV \textit{vs} 0.15 meV), while the degeneracy of $e_{g}$ orbitals is lifted due to larger difference of the orbital splitting from the Jahn-Teller and polar distortions ($-0.45$ eV \textit{vs} 0.13 eV). To confirm this idea, we calculate the PDOS of the structures obtained by linear interpolation between the strained bulk atomic structure without polarization ($c/a = 1.02$) parameterized by $\lambda=0$ and strained bulk atomic structure with out-of-plane polarization ($c/a=1.04$) parameterized by $\lambda=1$. Fig.~\ref{fig5} shows the PDOS with $\lambda=0, 0.4$, and $0.8$. For the all values of the $\lambda$, the splitting of the $e_{g}$ orbital is maintained due to the splitting from the elongated $c$-axis ($c/a > 1.02$) dominating the $e_{g}$ orbital splitting. In contrast, the sign change in the $t_{2g}$ orbital splitting is clearly seen for $\lambda=0$ and $\lambda=0.8$. In particular, for $\lambda = 0.4$, the $t_{2g}$ orbitals are closely degenerate while $e_{g}$ orbitals split about 0.3 eV, consistent with the experimental data. For the $\lambda = 0.4$, the polarization value reduced to about 42\% of the $\lambda=1$ structure, which may come from the finite thickness of BTO, suppressing the polarization.

\begin{figure*}[t]\vspace{-0pt}
	\includegraphics[width=0.85\textwidth]{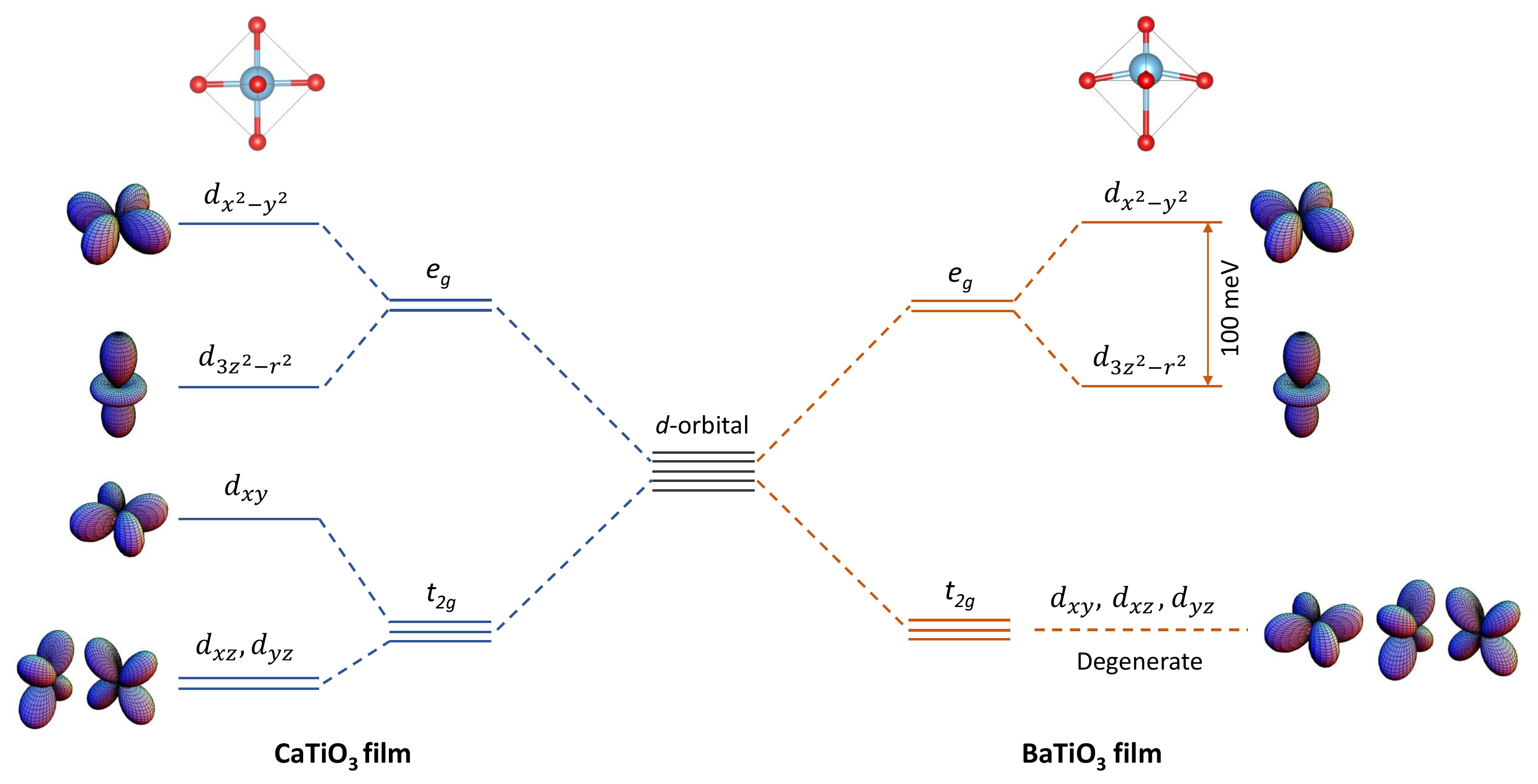}
	\caption{\label{fig6} Contrastive crystal field splitting of CTO films (left panel, without Ti-O displacements) and BTO films (right panel, withTi-O displacements).}
\end{figure*}

We compare the crystal-field-induced splitting between polar BTO without octahedral rotation and tilts and non-polar CaTiO$_{3}$ (CTO) with $a^{-}a^{-}c^{+}$ octahedral rotation and tilts commonly observed in perovskite oxides, illustrated in Fig.~\ref{fig6}. For TiO$_6$ octahedra under biaxial compressive strain, the orbital structure ($d_{xz}$/$d_{yz}$ $<$ $d_{xy}$ $<$ $d_{3z^2-r^2}$ $<$ $d_{x^2-y^2}$) in biaxial compressive CTO films is expected due to elongated Ti-O bonds along the $c$-axis. With combined tetragonal and polar distortions in BTO, a large orbital splitting between $e_{g}$ orbitals ($d_{3z^2-r^2}$ $<$ $d_{x^2-y^2}$) with an energy gap about 100 meV is observed from the dominated Jahn-Teller splitting from increased $c/a$ ratio, whereas $t_{2g}$ ($d_{xz}$/$d_{yz}$/$d_{xy}$) orbitals are nearly degenerate in the BTO thin film from the competition between polar and Jahn-Teller distortions. The unusual orbital structure we have observed in this work may provide a clue to understand the peculiar band splitting of BTO investigated by angle-resolved photoemission spectroscopy (ARPES) \cite{PRB-2017-Rodel}, by including the crystal field splitting from polar distortions. Our finding here provides an effective way to manipulate orbital degree of freedom by manipulating ferroelectric polarization which could be used to design exotic quantum states, such as metal-insulator transition, superconductivity, and colossal magneto-resistance.

\section{Conclusion}
In summary, we have synthesized high-quality BTO thin films on DSO substrates with a layer-by-layer growth by PLD and characterized their structural and electronic properties by RHEED, element-specific XAS/XLD, and DFT calculation. In sharp contrast to conventional crystal field splitting and orbital configuration ($d_{xz}$/$d_{yz}$ $<$ $d_{xy}$ $<$ $d_{3z^2-r^2}$ $<$ $d_{x^2-y^2}$) in elongated TiO$_6$ octahedra such as compressive CTO films, the XLD spectra reveals that the orbital structure in BTO films is unconventional: nearly degenerate $t_{2g}$ ($d_{xz}$/$d_{yz}$/$d_{xy}$) and split $e_{g}$ ($d_{3z^2-r^2}$ $<$ $d_{x^2-y^2}$ with a gap $\sim$ 100 meV) states. The first-principles DFT calculations show that this unexpected degenerate $t_{2g}$ orbitals are from the competition between polar and Jahn-Teller distortions in the non-centrosymmetric TiO$_6$ octhedra of BTO films. Our work could pave a way to design exotic quantum states (such as tunable multiferroic properties) by manipulating the orbital degree of freedom using the switchable ferroelectric polarizations.

\begin{acknowledgments}

We acknowledge insightful discussions with Darren C. Peets. This work is supported by the National Natural Science Foundation of China (Grant No. 11874058), the Pioneer Hundred Talents Program of the Chinese Academy of Sciences, the Ningbo 3315 Innovation Team, and the Ningbo Science and Technology Bureau (Grant No. 2018B10060). J. C. acknowledges support from the Gordon and Betty Moore Foundation EPiQS Initiative through Grant No.\ GBMF4534. This research used resources of the Advanced Light Source, which is a DOE Office of Science User Facility under contract No. DE-AC02-05CH11231. S. P. acknowledges support from the Institute for Basic Science in Korea (Grant No. IBS-R009-D1).
\end{acknowledgments}

%\newpage

%{\textbf{Supplementary}}

\end{document}